\documentclass{ifacconf}

\usepackage{graphicx}      
\usepackage{subcaption}
\usepackage{natbib}        
\usepackage{url}
\usepackage{amsmath}
\usepackage{amsfonts}
\usepackage{amssymb}
\usepackage{tikz}
\usepackage{siunitx}

\begin{document}
\begin{frontmatter}

\title{Onboard Model-based Prediction \\of Tram Braking Distance\thanksref{footnoteinfo}} 

\thanks[footnoteinfo]{The research was funded by Technology Agency of the Czech Republic within the program Epsilon, the project TH03010155.}

\author[LD]{Loi Do} 
\author[IH]{Ivo Herman}
\author[ZH]{Zden\v{e}k Hur\'ak} 

\address[LD]{Faculty of Electrical Engineering, Czech Technical University in Prague, CZ (e-mail: doloi@fel.cvut.cz).}
\address[IH]{Herman Electronika, Brno, CZ (e-mail: ivo.herman@herman.cz).}
\address[ZH]{Faculty of Electrical Engineering, Czech Technical University in Prague, CZ (e-mail: hurak@fel.cvut.cz)}

\begin{abstract}                
In this paper, we document design and a prototype implementation of a computational method for an onboard prediction of a breaking distance for a city rail vehicle---a tram. 
The method is based on an onboard simulation of tram braking dynamics. 
Inputs to this simulation are the data from a digital map and the estimated (current) position and speed, which are, in turn, estimated by combining a mathematical model of dynamics of a tram with the measurements from a GNSS/GPS receiver, an accelerometer and the data from a digital map. 
Experiments with real trams verify the functionality, but reliable identification of the key physical parameters turns out critically important. The proposed method provides the core functionality for a collision avoidance system based on vehicle-to-vehicle (V2V) communication. 
\end{abstract}

\begin{keyword}
Automatic control, optimization, real-time operations in transportation, Simulation, Braking distance prediction, Rail transport, Mathematical modeling
\end{keyword}

\end{frontmatter}

\section{Introduction}
\subsection{Motivation, short description}
In this paper, we present a computational method for prediction of a tram braking distance based on an onboard simulation of a mathematical model of tram braking dynamics. 
The proposed algorithm runs onboard the tram. 
It combines in real time the outputs from the mathematical model of longitudinal dynamics of the tram with the signals provided by a GNSS/GPS receiver, an inertial measurement unit (IMU), and the data read from a digital map.

The motivation for such development is the disturbingly high number of collisions of trams with other trams, other vehicles, and even pedestrians\footnote{Prague Public Transit, Co. Inc. has been registering well above one thousand collisions every year.}. 
Different collision scenarios call for different approaches to their solutions. 
In the research described in this paper, we restrict ourselves to collisions between vehicles. 
In particular, we consider tram-to-tram collisions. The reason for considering collisions between trams is that both participants in the (potential) collision are operated by a single organization/company, which makes coordinated collision avoidance schemes (almost) immediately realizable. 
The key technology for such coordination is wireless vehicle-to-vehicle (V2V) communication, through which the trams could exchange their position and speed estimates and predictions, thus truly distributed estimation/predictions could be realized.   

The parameters of the model of dynamics of the tram can be partially determined from the technical specifications provided by the producer and partially extracted from data measured using onboard sensors during experiments (grey box identification).
Although in this paper, we consider one particular tram brand and type, the proposed procedure can be applied to any (light) rail vehicle.

\subsection{State of the art}
In general, real time prediction of braking distance (or motion in general) of rail vehicles plays an important role in \textit{safety} application~\citep{lehner_reliable_2009}, \citep{gu_wireless_2013}, \citep{wu_position_2018} or \textit{energy optimization} \citep{lu_partial_2016}, \citep{keskin_energy_2016}. 
For safety applications involving suburban or freight trains, the prediction of braking distance need not be accurate due to relatively large gap distance (to the next train) and approximately constant parameters of the train dynamics.
Simple braking distance prediction models based only on constant parameters of heavy-weight trains their current speed and track characteristics have been published~\citep{ieee_ieee_2009}.
In cities with a dense tram network, however, the distance gap between two (rail or road) vehicles could go down to a few meters (stopping at tram stops or traffic lights), resulting in a higher probability of collisions. 
This imposes higher requirements on the accuracy of related onboard estimations and predictions which are, in turn, conditioned by the accurately identified physical parameters.
The parameters are mainly the weight of the vehicle, which is given by the actual number of passengers (the total weight of the tram can be some 50 \% higher from an empty tram), the coefficients characterizing the adhesion conditions (given by humidity, temperature), slope of the track (tramways). 
Not many works in the literature seem to be systematically addressing these issues.

One component of the full collision avoidance system is an estimator of the (distance) gap, that is, the distance between the rear bumper of the leading vehicle and the front bumper of the following vehicle. 
One class of these estimators is based on vision-based techniques~\citep{mukhtar_vehicle_2015}. 
An alternative followed in the broader project, within which we write this paper, is to use the V2V communication~\citep{xiang_research_2014},~\citep{abboud_interworking_2016} between the two vehicles and, effectively, realize an estimator of the distance gap in a distributed manner. 
Nonetheless, in this paper, we do not elaborate further on this topic of the distance gap estimation.

\subsection{Outline of the paper}
This paper is structured as follows. In Sec.~\ref{sec:HW}, we give some background information about the Tatra T3 tram. 
We also describe the instrumentation used for onboard measurements. 
Then in Sec.~\ref{sec:model} we describe the model of dynamics with the values of the physical parameters identified from real data obtained in experiments with trams and document the verification of the model.
In Sec.~\ref{sec:braking_distance}, we describe the use of the model for prediction of braking distance and describe in more detail estimation of the position and the speed of a tram. 
We also compare the proposed method with a simple equation-based braking distance prediction method.
Lastly, we give a conclusion in Sec.~\ref{sec:conclusion}.


\section{Experimental Setup}\label{sec:HW}
\subsection{Tatra T3 tram}
We focus on developing a model of dynamics of the Tatra T3 tram partially parametrized by data acquired onboard a tram during experiments.
With nearly 14 thousand produced trams during the period from 1960 to 1989, the T3 tram is one of the most produced tram cars in the world~\citep{mara_tatra_t3}.
In the Czech Republic, T3 trams (in several modernized versions) are still used nowadays and form a significant portion of public transport tram fleets in many cities.
Proportions of the tram and parameters relevant for the modeling are listed in Tab.~\ref{tb:t3_parameters}~\citep{linert_kolejova_vozidla}.


\begin{table}[!b]
\begin{center}
\captionsetup{width=.8\linewidth}
\caption{Parameters of the T3 tram.}\label{tb:t3_parameters}
\begin{tabular}{cccc}
Parameter 			& Notation			& Value 					\\\hline
Curb weight			& -					& $\approx\SI{16500}{\kilo\gram}$  	\\
Gross weight		& -					& $\approx\SI{27500}{\kilo\gram}$  	\\
Body length			& -					& $\SI{14000}{\milli\meter}$\\
Body height			& -					& $\SI{3060}{\milli\meter}$ \\
Body width			& -					& $\SI{1440}{\milli\meter}$\\
Wheel radius 		& $r$				& \SI{325}{\milli\meter} 	\\
Wheel mass			& $m_w$				& \SI{195}{\kilo\gram}		\\
Total motors power	& $P_\mathrm{max}$	& $4\times \SI{44}{\kilo\watt}$		\\
Maximum speed & - & $\SI{65}{\kilo\meter\per\hour}$ \\
\hline
\end{tabular}
\end{center}
\end{table}

\subsection{Instrumentation}
The instrumentation in the T3 tram measures only a few values, for instance, tram speed (computed from a wheel speed) or applied current to motor.
Also, we could not directly read the data during the experiments (lack of the CAN bus).
We, therefore, used external instrumentation to collect the measurements for the model identification.
We used GNSS reciever \emph{NEO‑M8P} by \emph{U-blox} to measure the position and the speed of the tram and inertial measurement unit (IMU) \emph{ADIS16465-1BMLZ} by \emph{Analog devices} to measure the acceleration. 
We used ready-to-use application/evaluation boards from the manufacturer to directly read the measurements from the sensors.
We set the sampling frequency of the GNSS/GPS reciever to \SI{1}{\hertz} and the sampling frequency of the IMU to \SI{2}{\kilo\hertz}.
We aligned the x-axis of IMU with the direction of the longitudinal motion of the tram.

\section{Model of dynamics}\label{sec:model}
\subsection{Identification of a model}

\begin{figure}[!tb]
	\begin{center}
	\includegraphics[width=8.4cm]{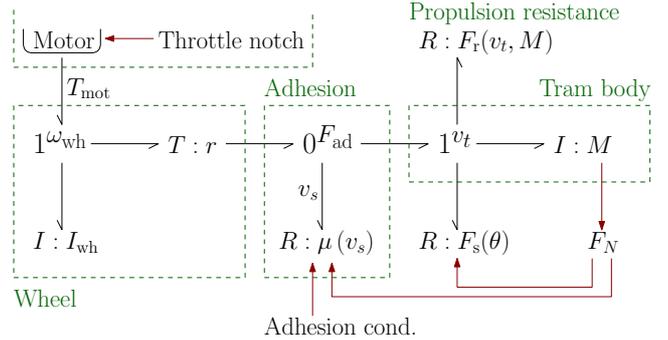}
	\caption{Bond graph of longitudinal motion dynamics.}\label{fig:lmd_bond_graph}
	\end{center}
\end{figure}

For the prediction of braking distance, it is sufficient to use the quarter model of dynamics describing only the longitudinal motion of a tram~\citep{sadr_predictive_2016}.
The quarter model of dynamics is given by equations:
\begin{subequations}\label{eq:LMD_equations}
\begin{align}
J_\mathrm{wh}\dot{\omega}_\mathrm{wh} &= \left(T_\mathrm{mot} - rF_\mathrm{ad}\right)\;, 
\\
M\dot{v_t} &= \left( F_\mathrm{ad} - F_\mathrm{r} - F_\mathrm{s}\right) \;, 
\end{align}
\end{subequations}
where $\omega_\mathrm{wh}$ is angular speed of the wheel, $J_\mathrm{wh}$ is moment of inertia of the wheel, $T_\mathrm{mot}$ is torque given from the motors, $r$ is radius of a wheel, $F_\mathrm{ad}$ is adhesion force, $v_t$ is longitudinal speed of the tram, $M$ is total weight of the tram, $F_r$ and $F_s$ are resistive forces.  
This model can be also described by a \textit{bond graph}~\citep{brown_engineering_2006} displayed in Fig.~\ref{fig:lmd_bond_graph}.
Approximating the wheel as a homogeneous disk, we can write its moment of inertia as:
\begin{equation}
J_\mathrm{wh} = 0.5m_wr^2\;.
\end{equation} 
The total weight of the tram $M$ is sum of the curb weight and weight of passengers in the tram.
%
The torque $T_\mathrm{mot}$ generated by the motors is proportional to the notch (position) of the control throttle.
In the T3, the throttle control has in total 15 notches: seven notches for acceleration, seven notches for deceleration, and one notch for idle.
We identified the traction characteristics experimentally as:
\begin{equation}
\tilde{T}_\mathrm{mot}(p) = 
\begin{cases}
1449p										&\text{for }p\geq0 \text{ and } \tilde{T}_\mathrm{mot}\omega_\mathrm{wh}	<  P_\mathrm{max} \;,\\
P_\mathrm{max}/\omega_\mathrm{wh}	&\text{for }p\geq0 \text{ and } \tilde{T}_\mathrm{mot}\omega_\mathrm{wh}	\geq  P_\mathrm{max} \;,\\
1176p										&\text{for }p<0 \;,\\
\end{cases}
\end{equation}
where $p \in \left\{-7, -6, \ldots, 0, \ldots, 6, 7\right\}$ is the notch.
The constants (\num{1449} and \num{1176}) were set to match experimentally measured maximal and minimal acceleration at the highest and lowest notch, respectively.
Braking torque is not restricted by the $P_\mathrm{max}$.
Also, we model the dynamics of the change of $\tilde{T}_\mathrm{mot}$ as first order system:
\begin{equation}
T_\mathrm{mot}(t) = 3e^{-3t}\tilde{T}_\mathrm{mot}(p) \;.
\end{equation} 

%
%
%
The adhesion force $F_\mathrm{ad}$ accounts for the transfer of wheel speed into the longitudinal motion of the tram body.
A physical explanation of the adhesion is given in \cite{park_modeling_2008}.
In general, the adhesion force $F_\mathrm{ad}$ is computed as a sum of adhesion force given by each traction wheel and the adhesion force is proportional to the wheel load.
However, since all wheels of T3 tram are connected to the traction motors and by assuming the uniform distribution of the tram weight on each wheel, we can directly write:
\begin{subequations}\label{eq:LMD_equations_v2}
\begin{align}
F_\mathrm{ad} &= \mu(v_s) F_N \;,	
\\
F_N &= Mg\;,
\\
v_s &= r\omega_\text{wh} - v_t \;, 	
\\
\mu(v_s) &= c_a e^{-a_av_s} - d_ae^{-b_av_s}\;,
\end{align}
\end{subequations}
where $g$ is gravitational acceleration.
Parameters $a_a$, $b_a$, $c_a$ and $d_a$ vary on track conditions~\citep{takaoka_disturbance_2000}.

\begin{figure}[!tb]
	\begin{center}
	\includegraphics[width=8.4cm]{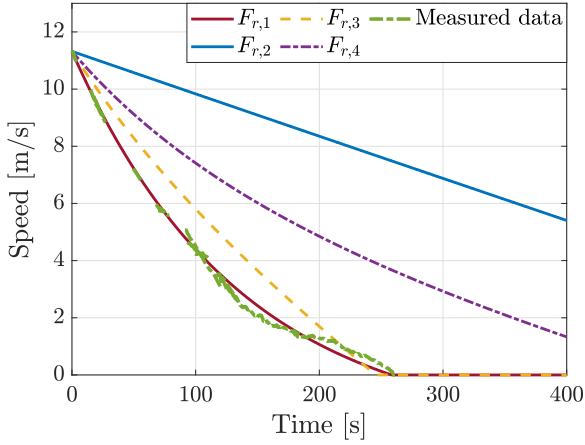}
	\caption{Comparison of identified parameters for propulsion resistance and parameters from literature.}\label{fig:propulsion_res_identified_compared}
	\end{center}
\end{figure}
Propulsion resistance $F_\mathrm{r}$ of rail vehicles (sum of rolling and air resistance) is typically calculated using an empirical equation in a form~\citep{hay_railroad_1982}:
\begin{equation}\label{eq:propulsion_equation}
F_\mathrm{r}(v_t) = A + Bv_t + Cv_t^2\;,
\end{equation}
where $A, B$ and $C$ are coefficients giving the dependence of propulsion resistance on characteristics of the rail vehicle such as weight, number of axles or front surface cross-sectional area.
To identify coefficients in Eq.~\eqref{eq:propulsion_equation} we first gather all measurements from experiments during which the tram decelerate only due to propulsion resistance (idle notch), see Fig.~\ref{fig:propulsion_res_identified_compared} (\textit{Measured data}).
Note, that we could not obtain longer continuous decay from higher speeds due to the restriction of the test track.
We can simulate the model and set the same conditions as in the experiments to find appropriate coefficients.
The weight of the tram during experiments was $\SI{17000}{\kilo\gram}$ (estimated from a number of people in the tram during the experiment). 
Since trams usually operate at lower speeds than other trains, we simplify the identification of Eq.~\eqref{eq:propulsion_equation} by neglecting quadratic dependence on $v_t$, thus setting $C=0$.
The identified propulsion resistance equation is:
\begin{equation}\label{eq:idenfified_propulsion}
F_\mathrm{r,1}(v_t, M) = 0.0147M + 125.83v_t\;.
\end{equation}
The reason why we do not use coefficients of Eq.~\eqref{eq:propulsion_equation} from the literature is that they are usually designed for rail vehicles which have significantly higher weight and also operate at higher speeds than trams.
For instance, using the following equations from the literature evaluated for the T3 tram yields:
\begin{subequations}\label{eq:propulsion_res}
\begin{align}
F_\mathrm{r,2}(v_t, M) &=  0.0147M + \num{2.18e-06}Mv_t^2\;, 
\\
F_\mathrm{r,3}(v_t, M) &= 520 + 0.0065M + 3.6v_t + 3.8880v_t^2\;,
\\
F_\mathrm{r,4}(v_t, M) &= 1.839 \sqrt{M} + 0.0036 Mv_t + 4.329v_t^2\;.
\end{align}
\end{subequations}
Propulsion resistance $F_\mathrm{r,2}$ was designed for passenger train on bogies~\citep{iwnicki_handbook_2006}, $F_\mathrm{r,3}$ for electric locomotive and $F_\mathrm{r,4}$ for suburban electric multiple unit train~\citep{rochard_review_2000}.
See the comparison in Fig.~\ref{fig:propulsion_res_identified_compared}.
We can see that the speed decay when using any of Eq.~\eqref{eq:propulsion_res} does not match with the measurements.

Lastly, in addition to propulsion resistance, the motion of a tram is also affected by the slope $\theta$ (longitudinal inclination) of track:
\begin{equation}
F_s(\theta) = F_N\sin\theta\;.
\end{equation}
Slope $\theta$ is positive for ascent and negative for the descent.

\subsection{Verification of a model}
\begin{figure}[!tb]
	\begin{center}
	\includegraphics[width=8.4cm]{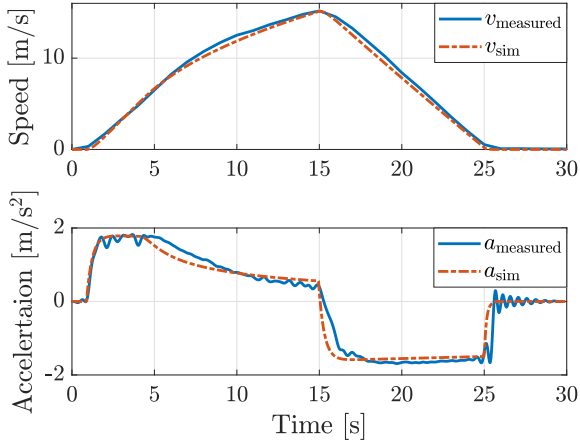}
	\caption{Comparison of the model and real measurements for given sequence of throttle notches: maximal and minimal acceleration.}\label{fig:verification_model_max_acc}
	\end{center}
\end{figure}

\begin{figure}[!tb]
	\begin{center}
	\includegraphics[width=8.4cm]{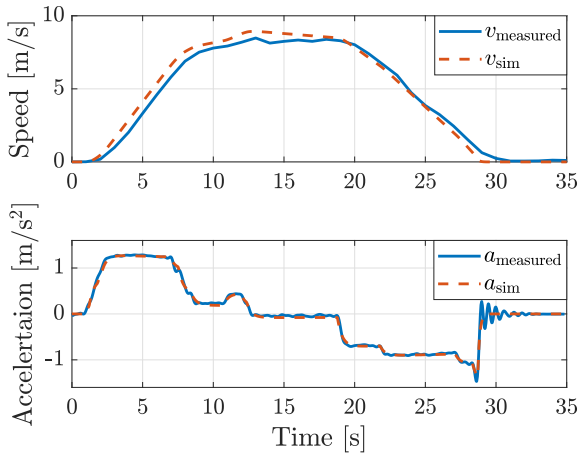}
	\caption{Comparison of the model and real measurements for given sequence of throttle notches.}\label{fig:verification_model}
	\end{center}
\end{figure}

To verify the model with experimentally obtained data, we first remove the noise and bias in the measured data from the accelerometer. 
The noise was contained in the measurements mainly due to vibrations onboard of a tram and sensor noise.
Since the longitudinal dynamics of a tram is relatively slow, we used the low-pass filter with the cut-off frequency \SI{2}{\hertz}.
We verify the model by setting the same sequence of the throttle notches, as in real experiments for two scenarios.
In the scenario in Fig.~\ref{fig:verification_model_max_acc}, the driver was instructed to set the maximal acceleration for a certain amount of time and then set the maximal deceleration.
We can see, that around time \SI{5}{\second}, $T_\mathrm{mot}\omega_\mathrm{wh} \geq P_\mathrm{max}$ which results in decline of $T_\mathrm{mot}$. 
The total actual and simulated traveled distance during the experiment is \SI{217}{\meter} and \SI{207}{\meter}, respectively.
The scenario in Fig.~\ref{fig:verification_model} has a more complicated sequence of throttle notches.
The total actual and simulated traveled distance during this scenario is \SI{158}{\meter} and \SI{164}{\meter}, respectively.
We can see that the modeled dynamics of the acceleration and the speed is similar to the experimental data.
Note that the discrepancies in total simulated and actual traveled distances are caused by a small error in generated acceleration, which is integrated twice.

The tram body at the moment of the stop exhibits oscillation due to the mechanics in the bogies.
This effect is not incorporated into the model since it does not affect the braking distance.

We will conduct another verification experiments with varying conditions (different track, adhesion conditions, and a total weight $M$ of the tram) by the end of the year 2019.

\section{Braking distance prediction}\label{sec:braking_distance}
To predict the braking distance using the model of dynamics, we first measure (or estimate) the speed $v_t$ of the tram.
The speed $v_t$ can be estimated, for instance, by fusing measurements from GNSS, IMU~\citep{bar-shalom_estimation_2008}, and possibly the wheel speed~\citep{ararat_robust_2017}.
The value of the speed $v_t$ is then used as an initial condition in the simulation of the model.
We can then simulate the model with a selected notch ($p\leq0$) and obtain the braking distance as the total traveled longitudinal distance from the start of the simulation until the tram reaches the speed $v_t=0$. 
Since the simulation of the proposed model is not computationally intensive, the simulation can be run periodically in real time onboard a tram, for instance, synchronously with the estimation of the speed $v_t$.

To get correct results, we need also to estimate all following time varying parameters:
\begin{itemize}
\item Adhesion conditions: Estimation of the adhesion conditions (coefficients of $\mu(v_s)$) is well covered by various methods~\citep{sadr_predictive_2016}, \citep{pichlik_locomotive_2018}.
\item Slope: The slope $\theta$ of the track can be retrieved from known absolute position and the digital map.
\item Weight: The weight $M$ could be estimated, for instance, from number of passengers in the tram counted by sensors at the tram entrances.
\end{itemize}
In the work which we present in this paper, we extensively focused only on estimation of the position and the speed.
We assume that the total weight and the adhesion conditions are known.

\subsection{Position and speed estimation}
\begin{figure}[!tb]
	\begin{center}
	\includegraphics[width=8.4cm]{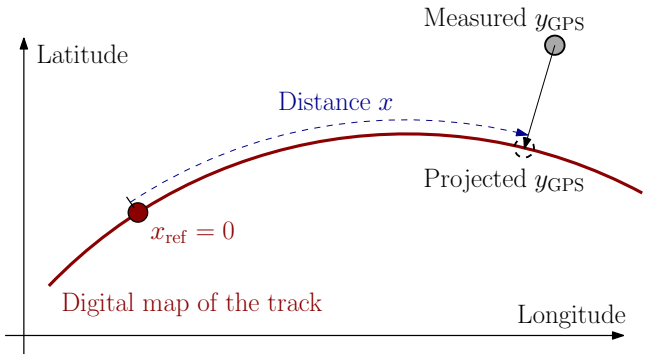}
	\caption{Illustraion of projection of measured point $y_\mathrm{GPS}$.}\label{fig:GPS_projection_onto_track}
	\end{center}
\end{figure}

To estimate the position and the speed of a tram, we use discrete Kalman Filter with measurements from IMU, GNSS/GPS receiver, and the digital map.
We first need to rewrite the equations~\eqref{eq:LMD_equations} and~\eqref{eq:LMD_equations_v2} into the discrete state-space representation.
Also, since we can not measure the motion of the wheel, nor the input from the driver in the T3 tram, we simplify the equations by assuming constant longitudinal acceleration:
\begin{equation}\label{eq:statespace_discrete}
\begin{bmatrix}
x_{k+1} \\
v_{k+1} \\
a_{k+1} \\
\end{bmatrix}
=
\begin{bmatrix}
1	&	\Delta t	&	0.5\Delta t^2	\\
0	&	1			&	\Delta t	\\
0	&	0			&	1	\\
\end{bmatrix}
\begin{bmatrix}
x_{k} \\
v_{k} \\
a_{k} \\
\end{bmatrix}\;,
\end{equation}
where $\Delta t$ is sample time, $x_k$, $v_k$, and $a_k$ are respectively the longitudinal traveled distance, the speed, and the acceleration.
The change in the acceleration $a_k$ is realized (when using the Kalman Filter)  through the measurements.
The output equation is:
\begin{equation}\label{eq:statespace_discrete_output}
y_k = 
\begin{bmatrix}
x_k\\
v_k\\
a_k\\
\end{bmatrix}\;.
\end{equation}
One problem with this description is that the GNSS receiver measures position as an absolute geographic coordinate $y_\mathrm{GPS}$, whereas the model output is longitudinal distance.
We resolve this problem by creating a mapping from $y_\mathrm{GPS}$ into longitudinal distance $x$ by use of a digital map of the track so the real measurement can be compared with model output.
The principle of the mapping is as follows: we first project the measured absolute geographic position $y_\mathrm{GPS}$ onto the track.
Then, we can compute traveled longitudinal distance $x$ from some selected reference point on the track $x=0$ (such as a depot or first tram stop) to the projected point.
See Fig.~\ref{fig:GPS_projection_onto_track} for an illustration.
The estimated absolute position, in combination with a digital map, also gives the current slope $\theta$.

\subsection{Comparison of braking distance prediction methods}
We now compare the proposed method for braking distance (model-based) prediction with the prediction using an equation (equation-based) derived from simple kinematics:
\begin{equation}\label{eq:braking_simple_eq}
d_\mathrm{br} = 0.5v_t^2a_\mathrm{dec}^{-1}		\;,
\end{equation}
where $d_\mathrm{br}$ is calculated braking distance and $a_\mathrm{dec}$ is selected braking deceleration. 
Such calculation of braking distance for rail vehicles has been published~\citep{ieee_ieee_2009} and used in several works~\citep{lu_partial_2016},~\citep{wu_position_2018} due to its simplicity.

\begin{figure}[!tb]
	\begin{center}
	\includegraphics[width=8.4cm]{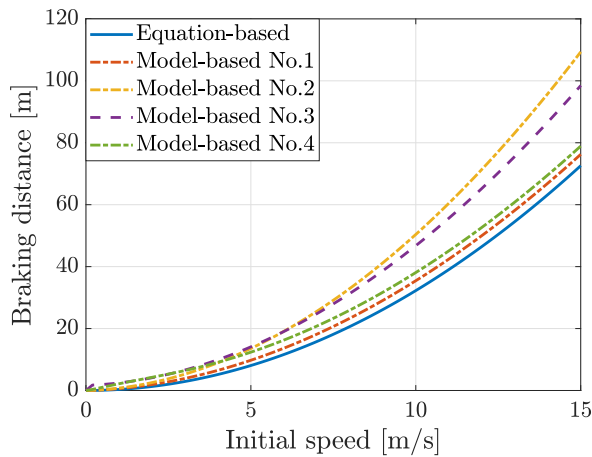}
	\caption{Comparison of equation-based and model-based prediction of braking distance.}\label{fig:braking_distance_comparison}
	\end{center}
\end{figure}

To get comparable results from equation-based and model-based methods, we set  maximal deceleration of equation-based prediction~\eqref{eq:braking_simple_eq} to $a_\mathrm{dec}=\SI{1.55}{\meter\per\second\squared}$ which approximately corresponds to maximal measured deceleration of empty tram with zero slope and dry adhesion conditions. 
We set such physical parameters in the simulation, see Fig.~\ref{fig:braking_distance_comparison}, simulation No.~1.
We can see that the equation-based model gives in general lower braking distance (at \SI{15}{\meter\per\second} lower by $\approx\SI{4}{\meter}$).
The difference is mainly to the fact that model-based calculation, unlike the equation-based calculation,  takes into account also the limited dynamics of deceleration.

For illustration, we also simulate the model with various parameters, see Tab.~\ref{tb:parameters_sim}.
Adhesion coefficients $[a_a,b_a,c_a,d_a]$ for dry conditions are: $[0.54,1.2,1,1]$ and for wet conditions: $[0.05,0.5,0.08,0.08]$~\citep{takaoka_disturbance_2000}.
From the simulation No.~2, we can observe, that even though a higher weight cause higher propulsion resistance, it also decrease maximal braking effort which results in significantly higher braking distance.
Braking distance is also significantly affected by the relatively small (negative) slope, as we can see from simulation No.~3.
Simulation No.~4 shows that worse adhesion conditions have a relatively higher effect on the braking distance at lower speeds ($\approx \SI{5}{\meter\per\second}$).

\begin{table}[!tb]
\begin{center}
\captionsetup{width=.8\linewidth}
\caption{Parameters of the simulations.}\label{tb:parameters_sim}
\begin{tabular}{ccccc}
Sim. No. 	& Notch $p$	& $M\,[\si{\kilo\gram}]$	& Slope $\theta\,[\si{\radian}]$			& Adh. cond.	\\
\hline
1					& $-7$		& $\num{17000}$		& 0							& Dry			\\
2					& $-7$ 		& $\num{25000}$		& 0							& Dry			\\
3					& $-7$		& $\num{17000}$		& ${-0.035}$				& Dry			\\
4 					& $-7$		& $\num{17000}$		& 0 						& Wet			\\
\hline
\end{tabular}
\end{center}
\end{table}

\section{Conclusion and future research}\label{sec:conclusion}
In this paper, we presented a method for prediction of tram braking distance based on real time simulation of a model of dynamics of a tram.
The braking distance prediction algorithm runs onboard a tram, taking as the inputs estimated longitudinal speed of a tram and key time-varying physical parameters affecting the braking distance such as the total weight of the tram, adhesion conditions, or the slope of the track.
We have shown, through various simulations, that predicted braking distance significantly depends on values of the physical parameters.
For safety applications such as collision avoidance, which is the ultimate goal of this research, the accurate prediction of braking distance is crucial for the correct detection of imminent collisions.

Besides the safety applications, within which we have written this paper, the model can be used in other applications such as energy optimization or slip control.
As a model of dynamics, we used a quarter model for a rail vehicle.
We identified the parameters of the model partially from the literature and partially from the data measured using onboard sensors during real experiments.
Also, by comparing the simulations of the model with real experiments, we verified its correctness.
Another verification experiments with different physical parameters will be done by the end of the year 2019.
The implemented model in Matlab and Simulink is downloadable at \url{https://www.mathworks.com/matlabcentral/fileexchange/73391}.

\begin{ack}
Significant help from Vít Obrusník with conducting the experiments is acknowledged. We would also like to express our gratitude to the engineers from Ostrava Public Transit Co. Inc. (Dopravní podnik Ostrava, a.s.) for enabling the experiments with trams. 
\end{ack}

\bibliography{vehicular_communication_networks,modeling_railroad_vehicles,slip_modeling_and_estimation,railway_collision_avoidance,railway_vehicle_position_estimation}             

\begin{thebibliography}{21}
\providecommand{\natexlab}[1]{#1}
\providecommand{\url}[1]{\texttt{#1}}
\providecommand{\urlprefix}{URL }
\expandafter\ifx\csname urlstyle\endcsname\relax
  \providecommand{\doi}[1]{doi:\discretionary{}{}{}#1}\else
  \providecommand{\doi}{doi:\discretionary{}{}{}\begingroup
  \urlstyle{rm}\Url}\fi

\bibitem[{Abboud et~al.(2016)Abboud, Omar, and
  Zhuang}]{abboud_interworking_2016}
Abboud, K., Omar, H.A., and Zhuang, W. (2016).
\newblock Interworking of {DSRC} and {Cellular} {Network} {Technologies} for
  {V}2x {Communications}: {A} {Survey}.
\newblock \emph{IEEE Transactions on Vehicular Technology}, 65(12), 9457--9470.
\newblock \doi{10.1109/TVT.2016.2591558}.
\newblock 00006.

\bibitem[{Ararat and S{\"o}ylemez(2017)}]{ararat_robust_2017}
Ararat, {\"O}. and S{\"o}ylemez, M.T. (2017).
\newblock Robust {Velocity} {Estimation} for {Railway} {Vehicles}.
\newblock \emph{IFAC-PapersOnLine}, 50(1), 5961--5966.
\newblock \doi{10.1016/j.ifacol.2017.08.1256}.

\bibitem[{Bar-Shalom et~al.(2008)Bar-Shalom, Li, and
  Kirubarajan}]{bar-shalom_estimation_2008}
Bar-Shalom, Y., Li, X.R., and Kirubarajan, T. (2008).
\newblock \emph{Estimation with {Applications} to {Tracking} and {Navigation}:
  {Theory} {Algorithms} and {Software}}.
\newblock Wiley-Interscience, 1st edition.

\bibitem[{Brown(2006)}]{brown_engineering_2006}
Brown, F.T. (2006).
\newblock \emph{Engineering {System} {Dynamics} : {A} {Unified}
  {Graph}-{Centered} {Approach}, {Second} {Edition}}.
\newblock CRC Press.
\newblock \doi{10.1201/b18080}.

\bibitem[{Gu et~al.(2013)Gu, Zuo, Su, Zhang, and Xu}]{gu_wireless_2013}
Gu, Q., Zuo, H., Su, R., Zhang, K., and Xu, H. (2013).
\newblock A {Wireless} {Subway} {Collision} {Avoidance} {System} {Based} on
  {Zigbee} {Networks}.
\newblock \emph{Journal of Communications}, 8(9), 561--565.
\newblock \doi{10.12720/jcm.8.9.561-565}.
\newblock
  \urlprefix\url{http://www.jocm.us/index.php?m=content&c=index&a=show&catid=121&id=574}.
\newblock 00000.

\bibitem[{Hay(1982)}]{hay_railroad_1982}
Hay, W.W. (1982).
\newblock \emph{Railroad {Engineering}}.
\newblock Wiley, New York, 2nd edition.

\bibitem[{{IEEE}(2009)}]{ieee_ieee_2009}
{IEEE} (2009).
\newblock {IEEE} {Guide} for the {Calculation} of {Braking} {Distances} for
  {Rail} {Transit} {Vehicles}.
\newblock \emph{IEEE Std 1698-2009}, C1--31.
\newblock \doi{10.1109/IEEESTD.2009.5332051}.

\bibitem[{Iwnicki(2006)}]{iwnicki_handbook_2006}
Iwnicki, S. (2006).
\newblock \emph{Handbook of {Railway} {Vehicle} {Dynamics}}.
\newblock CRC Press.
\newblock \doi{10.1201/9781420004892}.
\newblock \urlprefix\url{https://www.taylorfrancis.com/books/9781420004892}.

\bibitem[{Keskin and Karamancioglu(2016)}]{keskin_energy_2016}
Keskin, K. and Karamancioglu, A. (2016).
\newblock Energy {Efficient} {Motion} {Control} for a {Light} {Rail} {Vehicle}
  {Using} {The} {Big} {Bang} {Big} {Crunch} {Algorithm}.
\newblock \emph{IFAC-PapersOnLine}, 49(3), 442--446.
\newblock \doi{10.1016/j.ifacol.2016.07.824}.

\bibitem[{Lehner et~al.(2009)Lehner, De~Ponte~Müller, Strang, and
  Rico~García}]{lehner_reliable_2009}
Lehner, A., De~Ponte~Müller, F., Strang, T., and Rico~García, C. (2009).
\newblock Reliable vehicle-autarkic collision detection for railbound
  transportation.
\newblock In \emph{Proceedings of {ITS} 2009}. Stockholm, Schweden.
\newblock \urlprefix\url{http://elib.dlr.de/60056/}.
\newblock 00008.

\bibitem[{Linert et~al.(2005)Linert, Fojtík, and
  Mahel}]{linert_kolejova_vozidla}
Linert, S., Fojtík, P., and Mahel, I. (2005).
\newblock \emph{Kolejová vozidla pražské městské hromadné dopravy}.
\newblock Praha: Dopravní podnik hl. m. Prahy.

\bibitem[{Lu et~al.(2016)Lu, Wang, Weston, Chen, and Yang}]{lu_partial_2016}
Lu, S., Wang, M.Q., Weston, P., Chen, S., and Yang, J. (2016).
\newblock Partial {Train} {Speed} {Trajectory} {Optimization} {Using}
  {Mixed}-{Integer} {Linear} {Programming}.
\newblock \emph{IEEE Transactions on Intelligent Transportation Systems},
  17(10), 2911--2920.
\newblock \doi{10.1109/TITS.2016.2535399}.

\bibitem[{Mara(2001)}]{mara_tatra_t3}
Mara, R. (2001).
\newblock \emph{Tatra T3 1960-2000 : 40 let tramvají Tatra T3}.
\newblock K-Report.

\bibitem[{Mukhtar et~al.(2015)Mukhtar, Xia, and Tang}]{mukhtar_vehicle_2015}
Mukhtar, A., Xia, L., and Tang, T.B. (2015).
\newblock Vehicle {Detection} {Techniques} for {Collision} {Avoidance}
  {Systems}: {A} {Review} - {IEEE} {Journals} \& {Magazine}.
\newblock \doi{10.1109/TITS.2015.2409109}.

\bibitem[{Park et~al.(2008)Park, Kim, Choi, and Yamazaki}]{park_modeling_2008}
Park, S.H., Kim, J.S., Choi, J.J., and Yamazaki, H.o. (2008).
\newblock Modeling and control of adhesion force in railway rolling stocks.
\newblock \emph{IEEE Control Systems Magazine}, 28(5), 44--58.
\newblock \doi{10.1109/MCS.2008.927334}.

\bibitem[{Pichlik and Zdenek(2018)}]{pichlik_locomotive_2018}
Pichlik, P. and Zdenek, J. (2018).
\newblock Locomotive {Wheel} {Slip} {Control} {Method} {Based} on an
  {Unscented} {Kalman} {Filter}.
\newblock \emph{IEEE Transactions on Vehicular Technology}, 67(7), 5730--5739.
\newblock \doi{10.1109/TVT.2018.2808379}.

\bibitem[{Rochard and Schmid(2000)}]{rochard_review_2000}
Rochard, B.P. and Schmid, F. (2000).
\newblock A review of methods to measure and calculate train resistances.
\newblock \emph{Proceedings of the Institution of Mechanical Engineers, Part F:
  Journal of Rail and Rapid Transit}, 214(4), 185--199.
\newblock \doi{10.1243/0954409001531306}.

\bibitem[{Sadr et~al.(2016)Sadr, Khaburi, and
  Rodríguez}]{sadr_predictive_2016}
Sadr, S., Khaburi, D.A., and Rodríguez, J. (2016).
\newblock Predictive {Slip} {Control} for {Electrical} {Trains}.
\newblock \emph{IEEE Transactions on Industrial Electronics}, 63(6),
  3446--3457.
\newblock \doi{10.1109/TIE.2016.2543180}.

\bibitem[{Takaoka and Kawamura(2000)}]{takaoka_disturbance_2000}
Takaoka, Y. and Kawamura, A. (2000).
\newblock Disturbance observer based adhesion control for {Shinkansen}.
\newblock In \emph{6th {International} {Workshop} on {Advanced} {Motion}
  {Control}. {Proceedings} ({Cat}. {No}.00TH8494)}, 169--174.
\newblock \doi{10.1109/AMC.2000.862851}.

\bibitem[{Wu et~al.(2018)Wu, Wei, Weng, and Deng}]{wu_position_2018}
Wu, Y., Wei, Z., Weng, J., and Deng, R.H. (2018).
\newblock Position {Manipulation} {Attacks} to {Balise}-{Based} {Train}
  {Automatic} {Stop} {Control}.
\newblock \emph{IEEE Transactions on Vehicular Technology}, 67(6), 5287--5301.
\newblock \doi{10.1109/TVT.2018.2802444}.

\bibitem[{Xiang et~al.(2014)Xiang, Qin, and Xiang}]{xiang_research_2014}
Xiang, X., Qin, W., and Xiang, B. (2014).
\newblock Research on a {DSRC}-{Based} {Rear}-{End} {Collision} {Warning}
  {Model}.
\newblock \emph{IEEE Transactions on Intelligent Transportation Systems},
  15(3), 1054--1065.
\newblock \doi{10.1109/TITS.2013.2293771}.

\end{thebibliography}
\end{document}